\documentstyle[preprint,pre,aps]{revtex}

\tightenlines

\begin{document}

\draft

\title{Finite-Size Scaling in Two-dimensional \\ Continuum Percolation Models}

\author{Van Lien Nguyen\thanks{E-mail: nvlien@bohr.ac.vn}}
\address{
Institute of Physics, P.O. Box 429, Bo Ho, Hanoi 10000 Vietnam}
\author{Enrique Canessa\thanks{E-mail: canessae@ictp.trieste.it}}
\address{
The Abdus Salam International Centre for Theoretical Physics \\
P.O. Box 586, 34100 Trieste, Italy}

%\date{\today}

\maketitle

\begin{abstract}

We test the universal finite-size scaling of the cluster mass order
parameter in two-dimensional (2D) isotropic and directed continuum  
percolation models below the percolation threshold by computer 
simulations.  We found that the simulation data in the 2D continuum 
models obey the same scaling expression of mass $M$ to sample size 
$L$ as generally accepted for isotropic lattice problems, but with
a positive sign of the slope in the ln-ln plot of $M$ versus $L$.
Another interesting aspect of the finite-size 2D models is also 
suggested by plotting the normalized mass in 2D continuum and 
lattice bond percolation models, versus an effective percolation 
parameter, independently of the system structure ({\it i.e.}, lattice 
or continuum) and of the possible directions allowed for percolation 
({\it i.e.}, isotropic or directed) in regions close to the percolation
thresholds.   Our study is the first attempt to map the scaling 
behaviour of the mass for both lattice and continuum model systems
into one curve.

\end{abstract}

\vspace{0.7 cm}

\baselineskip=22pt

Isotropic percolation (IP) and directed percolation (DP) in Bravais
lattices and continuum percolation models have received considerable 
attention over the years \cite{Stau94,Shkl84,Hosh97,Por97}.  Such 
interests in (either lattice or continuum) percolation models, and
the crossover from IP to DP \cite{Frey94,Lien95,Lien98,Fro97}, stem
from theoretical reasons to understand a vast range of physical 
phenomena, including hopping conduction in strong electric fields
\cite{Lien81}, self-organized criticality \cite{Pacz96}, directed
polymers \cite{Lebe95}, porous silicon materials \cite{Yeh98}, {\em etc}.  
Most investigations until now
have mainly focused on the properties of clusters at the percolation
threshold for system sizes $L \rightarrow \infty$.  In this limit, IP
and DP critical clusters have been characterized by the correlation
length exponent $\nu$ and the fractal dimension given by 
$D_{f}=\ln M_{c}(L)/\ln L$, where $L\rightarrow \infty$ and $M_{c}(L)$
is the total number of sites belonging to the critical cluster 
({\em i.e.}, the cluster mass) at percolation.  Similar analysis 
have been used in a variety of different branches of physics 
\cite{Stau94}.  The shape and size of clusters with critical 
behaviour have been investigated, for instance, in Ref.\cite{Domb77}.  

For large, but still finite percolating systems prior to the 
percolation threshold, scaling assumptions have been studied 
based on the renormalization group treatment \cite{Stau94}.  Universal 
finite-size scaling below the percolation threshold $p_{c}$ has been 
discussed in a number of recent works \cite{Mar84,Nijs79,Aha97}.  It has
been suggested that the system order parameters $\Theta$ should scale 
with the finite system size $L$ such that 
\begin{equation}\label{eq:sca}
\Theta \sim L^{- \beta /\nu } F[ \lambda L^{1/\nu } ] \;\;\; ,
\end{equation}
where $\lambda = p_{c} - p$ is the separation from the critical point 
and, for 2D systems, $\beta$ and $\nu$ have the supposedly exact 
values 5/36 and 4/3, respectively \cite{Nijs79}.
For the order parameter of the largest cluster mass $M(L)$ in the
disordered phase (below percolation threshold), the plots
ln~$M(L)L^{\beta/\nu}$ versus ln~$L^{1/\nu}$ for the square, triangular
and cubic lattices of \cite{Mar84}, support the scaling of 
Eq.(\ref{eq:sca}) with a negative slope.
At the threshold,  such results has been shown to be sensitive to
system shape, boundary conditions, and other factors \cite{Aha97}.
However, this universality class phenomenon for the mass variations of
ramified clusters along the percolation processes has remain unexplored
in the case of continuum percolation models.

In this work we test the universal finite-size scaling of the cluster 
mass order parameter in 2D continuum percolation below percolation
threshold.  By investigating the behaviour of the cluster 
mass in these systems, we hope to shed light on the properties
of the internal structure of these percolation clusters below criticality.
A detailed knowledge of the properties of the complex 
topology of the available percolating paths would be useful
to study transport processes near the percolation threshold \cite{Por97}. 

We have carried out simulations using standard techniques.  In the continuum
IP and DP models, the coordinates $\{x_{i},y_{i}\}$ of $N$ sites are
generated at random in a square box of size $L = N^{1/2}$.  The simulation
length unit is chosen such that the density of sites, namely $n$, in the
box is always equal to unity regardless of the total number of sites $N$. 
Percolation is checked over sites from one (the left) edge to the opposite
(the right) edge of the simulation square ($x$-axis). 

In the continuum IP model we solve the circle problem \cite{Shkl84}, where 
all directions are equivalent and a connected bond can be established between
the $i$-site to the $j$-site if the following condition is satisfied
\begin{equation}\label{eq:1}
\sqrt[2]{ (x_{j}-x_{i})^{2}+(y_{j}-y_{i})^{2} }  \le r \;\;\; . 
\end{equation}
The radius $r$ is chosen to check the percolation process.  In the present
study, the IP critical radius is found to be for example $r^{(I)}_{c}=1.1992$
for a sample size $N=12800$ and $r^{(I)}_{c}=1.1991$ for $N=25600$. 
The reported simulation data include all digits which do not change with 
increasing number of realizations used for the averaging.  We estimate
error bars to be of the order $\pm 0.001$.
 
In the continuum DP model the connected bonds can be established from site 
$i$ to the forward sites $j$ only.  Therefore in addition to Eq.(\ref{eq:1}),
there is also the additional condition $x_{j}>x_{i}$ ({\em i.e.}, the
forward semi-circle problem \cite{Lien95,Lien98}).  The DP radius is measured
and found to be, for example, $r^{(D)}_{c}=1.301$ for a sample size $N=12800$
and $r^{(D)}_{c}=1.311$ for $N=25600$.  Note that the dimensionless 
quantities $r_{c}$ are here given in units of $n^{-1/2}$.  The different 
sources of the finite-size effect for the DP are discussed in \cite{Lien95}. 

In the standard percolation problems all characteristics quantities
($\nu$, $D_{f}$ ...) are measured at criticality: {\em i.e.}, at
$r=r^{(I)}_{c}$ (or $r^{(D)}_{c}$).  In the following we shall investigate
the finite scaling behaviour of the cluster mass $M(r)$ when $r$
of Eq.(\ref{eq:1}) for IP (plus the constrain $x_{j}>x_{i}$ for DP)  are
smaller than $r^{(I)}_{c}$ (or $r^{(D)}_{c}$).  

To check the scaling of Eq.(\ref{eq:sca}), the mass $M(r)$ of the largest 
cluster has been measured for different distances smaller than $r_{c}$ in
systems of various sizes:  $N$=400, 800, 1600, 3200, 6400, 12800 and 25600.
For given $N$ and $r$, the averaging mass $<M(r)>$ is taken over a number
of random realizations, which decreases from 12800 to 200 as $N$ increases
from 400 to 25600, respectively.

Figure 1 shows the finite-size scaling ln-ln plot of the measured mass
$<M(r)>$ times $L^{\beta / \nu}$ for both IP and DP continuum problems,
against the system size $L$ to the $1/\nu$ power.  As indicated, different
curves are for different radius $r<r_{c}$ (in units of $n^{-1/2}$).  Here,
we use $\nu=4/3$ for 2D IP and $\nu =\nu_{||}=1.67$ (the parallel 
exponent \cite{Lien95}) for the 2D DP model.  In both
cases, the used value of $\beta$ is  5/36 \cite{Mar84}.  Note that the
magnitude of $\beta$ does not affect the qualitative behaviour of Fig.1,
but leads to a change in the slope magnitude only.  It is clear that for
 all values of $r$ in this study and for both IP and DP continuum
models, the simulation points lie well along straight lines.
 
Thus, our simulation data confirm the functional form of Eq.(\ref{eq:sca}) 
suggested for isotropic lattices prior to the percolation threshold
\cite{Mar84}.  In the continuum cases, however, the slopes in the ln-ln 
plots shown in Fig.1 are found to be positive contrary to the negative 
ones observed in isotropic 2D lattices in the disordered phase below the
percolation thershold.  This allows us to assume that the finite-size 
scaling of Eq.(\ref{eq:sca}) is so universal that it is valid for any
percolation (lattice or continuum) models, regardless of isotropicness
or directedness. For 2D continuum percolation models (both IP and DP) we 
found positive slopes such that, for low $L$, the mass $M$ of the
finite largest cluster increases with increasing $L$.  It is feasible
to believe that such an increase should become saturated and $M$ will be 
essentially independent of $L$ when $L \rightarrow \infty$ according to
the general statement of finite-size scaling theory (see, {\it i.e.} 
\cite{Stau94} page 71).  We also note here the suggestion that the
percolation phenomena (even two- three-dimensional) are essentially 
formulated by one-dimensional percolation paths \cite{Hun90,Zvy95}.
It is likely that for different distances separation $r_{c} - r$ the 
satured mass should be different and might relate to different 
magnitudes $\epsilon$ of the slopes shown in Fig.1.  Hence the nature 
of the investigated model is reflected in the sign (and magnitude) of
the slopes for the scaling.  The present phenomenological scaling
behaviour qualitatively agrees with the results of the 2D Ising model
for $p<p_{c}$ \cite{Mar84}.

In both the IP and DP continuum models, we also measure the normalized 
mass
\begin{equation}\label{eq:2}
f(r) \equiv \ln <M(r)>/\ln L^{D_{f}} = \ln <M(r)>/\ln <M_{c}(L)> \;\;\; ,
\end{equation}
where $M(r)$, as above, corresponds to the mass of the largest finite
cluster when $r$ is smaller than $r^{(I),(D)}_{c}$.  
We attempt to show next that the normalized mass of Eq.(\ref{eq:2}) 
posseses an universal behaviour not only for both IP and DP
continuum models, but also for 2D bond lattice problem with a
suitable relation between the radius $r$ in the continuum problem and
the percolation probability in the bond problem.

To check for universality in 2D isotropic lattice percolation, we use a
standard 2D bond percolation problem in a square lattice.  For a given
value of parameter $p$ (where $1-p$ corresponds to the standard
percolation probability), there is a bond between the site $i$ to the
neighbor site $j$ if the random probability $p_{ij}$ is not lower than
$p$: {\em i.e.}, $p_{ij} \ge p$.  This definition of $p$ has been 
adopted for convenience since it is simpler to formulate a 
relation between the 2D percolation models under consideration.
The measured critical value is $p_{c}=0.504$ for a system of 
$120\times 120$ (averaged over 400 realizations).  Clearly, if 
the chosen value of $p$ with the percolation condition 
$p_{ij} \ge p$ is much greater than $p_{c}$, then the connected
bonds cannot percolate through the system since the ramified forming 
clusters are small in comparison with the sample sizes.  This situation
is similar to the case of continuum model systems with $r < r_{c}$ in 
Eq.(\ref{eq:1}).   Again we measure the cluster mass $<M(p)>$ for 
various values $p>p_{c}$ ({\it i.e.}, percolation threshold).

In order to map the scaling behaviour of the mass for both lattice and
continuum model systems into one curve in the regions close to the 
percolation threshold, we propose the
relation
\begin{equation}\label{eq:propo}
p \propto exp \; (-r/r_{c}) \;\;\; , \;\;\; r \rightarrow r_{c}  \;\;\; .
\end{equation}
This leads to the relation between the parameter $r^{*} \equiv r/r_{c}$
and the ratio $p^{*}\equiv p/p_{c}$ in regions close percolation thresholds:
\begin{equation}\label{eq:3}
p^{*} =  exp \; (1-r^{*}) \;\;\; , \;\;\;   r^{*} \rightarrow 1.
\end{equation}
For a given value of $r^{*}$, from Eq.(\ref{eq:3}) we have the
corresponding value of $p$ for the lattice problem.  Though this equation
should be valid only in regions close to the percolation thresholds 
($r{*} \rightarrow 1$ and, correspondingly, $p{*} \rightarrow 1$), our 
simulations are carried out for a large range of value of 
$p$ from 0.919 to 0.515 which correspond to  $r^{*}$ between 0.4 to
0.98.  We thus measure $<M($p$)>$, and the normalized mass of 
Eq.(\ref{eq:2}), for the value of $p$ corresponding to the simulation
value of $r^{*}$ in the continuum problems. 

As mentioned above, we have carried out large scale simulations and 
measure $f(r)$ for isotropic and directed continuum percolation clusters
for different $r$ as well as the normalized mass in the isotropic lattice 
model at different values of the normalized radii $r^{*}$, namely
\begin{equation}\label{eq:5}
g(r^{*})\equiv \ln <M(p^{*})>/\ln <M_{c}(1)> \;\;\; ,
\end{equation}
with $p^{*}$ related to $r^{*}$ via Eq.(\ref{eq:3}).
We note that simulations of continuum (or random) systems are much more
involved than those for lattices \cite{Shkl84}.  They need a larger 
amount of computer memory (to storage the site coordinates), and of 
CPU time (to calculate the distances between sites) \cite{Hosh97}. 

Figure 2 shows the normalized masses $f$ of Eq.(\ref{eq:2}) and $g$ 
of Eq.(\ref{eq:5}) as a function of the common variable of the reduced
radius $r^{*}$ and for large samples sizes $L$ as indicated. 
Surprisingly, these two functionals of $f(r^{*})$ (for both 2D IP and DP
continuum models) and of $g(r^{*})$ (for 2D bond square lattice problem)
fall in a common form over the large range of the variable 
$r^{*} \ge 0.4$ (though Eq.(\ref{eq:5}) is suggested to be only valid
in the region $r^{*} \rightarrow 1$).  From
statistics in several samples, we estimate the errors of our measurements
to be of the order of $5\%$.  Greater variations at low $r$ are a
consequence of the very small clusters (or even largest) with very few
sites only, so giving rise to bigger relative fluctuations on the 
average $<M>$ in this limit. 

These measurements of Fig.2 suggest a novel aspect of the finite-size regime 
independently of the system structure ({\it i.e.}, lattice or continuum)
and of the possible direction(s) allowed for percolation ({\it i.e.},
isotropic or directed) at least in regions close to the percolation 
thresholds.  We have verified that the same macroscopic 
behaviour is also approximatively valid for the normalized ratio 
$\ln M(r^{*})/\ln L^{D_{f}}$ for a single realization in (say, infinitively)
large lattices.  If, for example, in continuum models $r<{\bar r}_{ij}$, 
({\em i.e.}, $r$ is smaller than the average distance between sites), then
the total number of sites percolating is also small and there is only a 
weakly increase of $M$ with increasing $r$ in this region.  This is so 
because the originating cluster can only flow to reach a few sites in the
neighborhood close to it.  When $r$ is close to $r_{c}$, say
$(r_{c}-r)/r_{c} \sim 0.1$ (see, $r^{*} > 0.9$ in Fig.2), the cluster mass 
in IP and DP increases
sharply to fit the value (of the order) of the total number of sites in
the system.  For the critical value $r^{*}=1$ the cluster flows through
the whole system and reaches the right-edge of the simulation box.  The
size of the critical fractal cluster is then equal to the system size. 

In summary, we have dealt with the critical properties of 2D isotropic 
and directed percolation, in particular with the differences between
continuum and lattice versions of this process.  We have tested the 
finite-size scaling of the cluster mass order parameter in 2D continuum
models below the percolation threshold.  The nature of the system 
structure, {\em i.e.} lattice or continuum, relates to the sign and
magnitude of the slope for the functional dependence of $<M>$ in 
Eq.(\ref{eq:sca}) and shown in Fig.1.  Our findings also show that the
total number of sites in the largest cluster (cluster mass at variable
simulation parameters close to percolation threshold), becomes 
independent of the model structure
and of the directions allowed to percolate when normalized with respect
to the total number of sites in the critical 
cluster at the percolation threshold.  The differences between the processes
of IP and DP are only in the changes of the critical exponents $\nu$, the
fractal dimension $D_{f}$ and
the percolation threshold shift \cite{Stau94,Shkl84,Lien95}.  Surprisingly, 
the cluster mass (prior and up to the percolation threshold) fits into the 
same form as shown in Fig.2.  Though we are unable to justify what observed
in this figure, we believe that this universality should hold in regions
close to critical points. And, since for the lattice models both the bond 
and site problems belong to the same universality class \cite{Stau94},
the present universality should be observed for the site
problem and for different 2D lattices.  Our study is the first attempt to
map the scaling behaviour of the mass for both lattice and
continuum model systems into one curve.

In order to understand better these phenomena, it would be interesting to
study similar finite-size behaviour in more complex lattice structures, 
{\em i.e.}, anisotropic \cite{Lien98}, aperiodic and in higher dimensions 
\cite{Gala97}.  We hope that as a first attempt in mapping together two
percolation models, lattice and continuum, our present results
will stimulate more discussions on this topic.

One of the authors (VLN) is indebted to Prof A. Aharony for helpful
discussions.

\newpage

\begin{figure}
\caption{
$\ln - \ln$ plot of the measured mass $<M(r)>$ times $L^{\beta / \nu}$ 
in the 2D continuum models versus the system size $L$ to the 
$1/\nu$ power for isotropic percolation (IP: open symbols) and directed
percolation (DP: full symbols) as a function of model radius $r<r_{c}$.
$\epsilon$ is the magnitude of each slope.}
\end{figure}

\begin{figure}
\caption{
Normalized mass $f(r^{*})$ of Eq.(\ref{eq:2}) in the continuum models and 
normalized mass $g(r^{*})$ of Eq.(\ref{eq:5})
in the lattice problem (using condition in Eq.(\ref{eq:3})) as a function
of the parameter $r^{*}$ in Eq.(\ref{eq:1}). }
\end{figure}

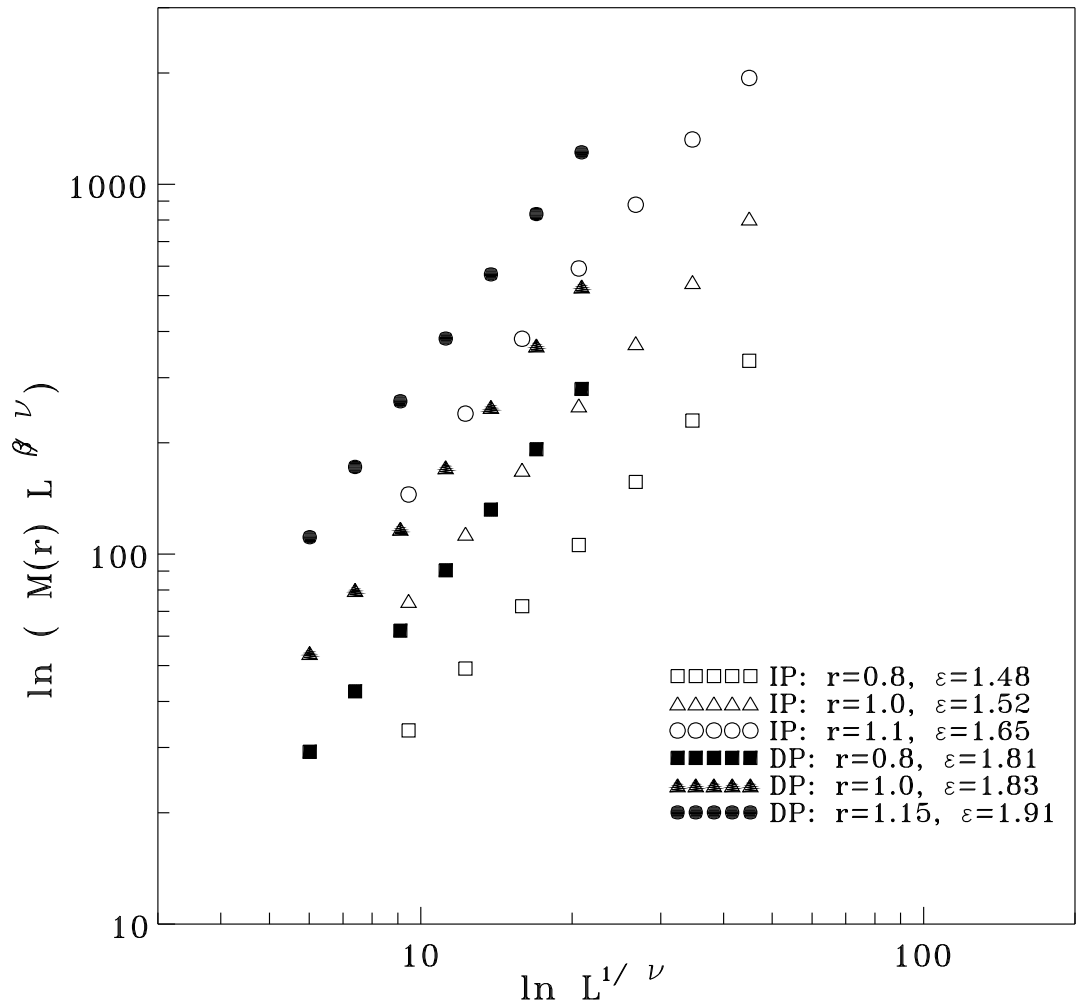
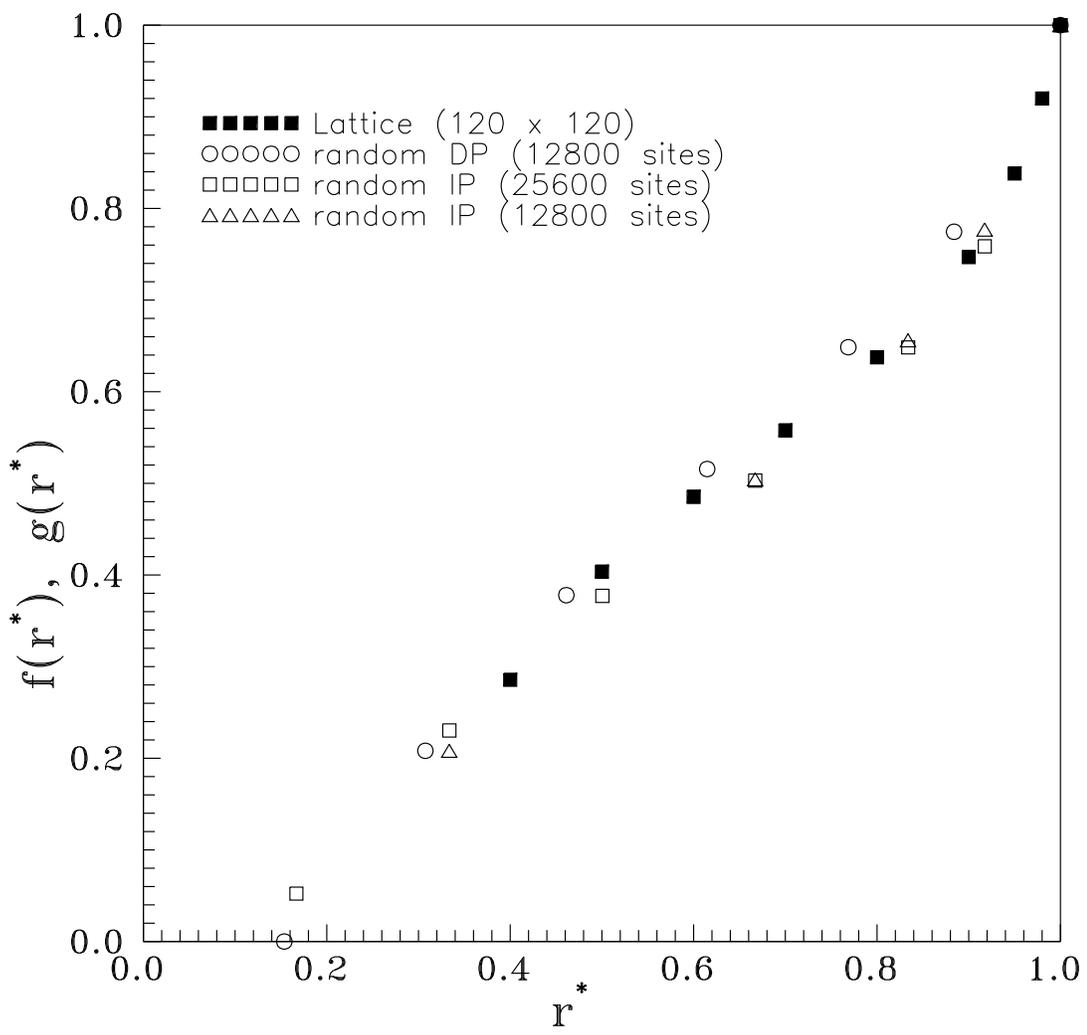
\end{document}